\documentclass[]{IEEEtran}
\usepackage{psfig}
\begin{document}

\title{The Rich-Club Phenomenon In The Internet Topology}
\author{Shi Zhou and Ra\'ul J. Mondrag\'on
\thanks{Manuscript received November 29, 2002. The associate editor coordinating the
review of this paper and approving it for publication was Prof. J.
Choe. This work is supported by the U.K. Engineering and Physical
Sciences Research Council (EPSRC) under Grant GR-R30136-01.}
\thanks{S. Zhou and R. J. Mondrag\'on are with the Department of Electronic Engineering,
Queen Mary, University of London, Mile End Road, London, E1 4NS,
United Kingdom (e-mail: shi.zhou@elec.qmul.ac.uk;
r.j.mondragon@elec.qmul.ac.uk). } }

\markboth{IEEE Communications Letters}{IEEE Communications
Letters}

\maketitle

\begin{abstract}
We show that the Internet topology at the Autonomous System (AS)
level has a rich--club phenomenon. The rich nodes, which are a
small number of nodes with large numbers of links, are very well
connected to each other. The rich--club is a core tier that we
measured using  the rich--club connectivity and the node--node
link distribution. We obtained this core tier without any
heuristic assumption between the ASes. The rich--club phenomenon
is a simple qualitative way to differentiate between power law
topologies and provides a criterion for new network models. To
show this, we compared the measured rich--club of the AS graph
with networks obtained using the Barab\'asi--Albert (BA)
scale--free network model, the Fitness BA model and the Inet--3.0
model.
\end{abstract}

\begin{keywords}
Internet, topology, modeling, networks.
\end{keywords}

\section{Introduction}
\PARstart{T}{he} analysis of the Internet topology at the
Autonomous System (AS) level by Faloutsos~{\sl et~al}
\cite{Faloutsos99} showed that the probability that a node has $k$
links has a power--law tail for large $k$, following $P(k)\propto
k^{-y}$, $y=2.22$. Subramanian~{\sl et~al}~\cite{Subramanian02},
using a heuristic argument based on the commercial relationship
between ASes, found that the Internet has a tier structure. Tier 1
consists of a  ``core'' of ASes which are well connected to each
other. There have been some attempts to model the network using
{\sl Transit--Stub} and {\sl Tiers} generators.
Tangmunarunkit~{\sl et~al} \cite{Tangmunarunkit02} showed that the
degree distributions produced by these structure-based generators
are not power--laws. Their results are based on qualitative
metrics and they recognized that there is a need for further
studies to characterize network topologies. There exist network
models that produce power--law networks, e.g. the Barab\'asi and
Albert (BA) scale--free model \cite{Barabasi99a} and the Inet--3.0
model \cite{Winick02} to mention just two of them.

We address in this letter the following questions: Can we
characterize the core tier of the AS without making any heuristic
assumptions? Does the power--law network generators produce a tier
structure similar to the one measured in the Internet? To answer
these questions we introduce the rich--club phenomenon as a
quantitative way to characterize a core tier without making any
heuristic assumption on the network elements interaction.

One of the main properties of power--law networks is that a small
number of nodes have large numbers of links, we call these nodes,
rich nodes. In this letter we show that the AS graph shows a
rich--club, i.e. a core tier. The members of the club tend to be
very well connected between each other, they create a  tight group
where if two members of the club do not share a link, it is very
likely that they share a common node that can link them, that is
the average hop distance is between one and two. Also, we have
compared the rich--club measured in the AS graph with the one
produced by the BA model, the Fitness BA model \cite{Bianconi00}
and the Inet--3.0 model, where the synthetic networks are created
to model the AS graph. Our results show that the BA and Fitness BA
model do not create a rich--club. The Inet--3.0 model creates a
rich--club but with a deficit in the number of core--links. Notice
that in this letter, we are not trying to characterize all the
existing power--law network generators, but to show that it is
possible to distinguish between them by studying the properties of
the rich--club.

\section{The AS Graph And Its Models}

\begin{table*}
\caption{Parameters of the four networks} \label{table_AS}
\centering
\renewcommand{\arraystretch}{1.2}
\begin{tabular}{c c c c c}
\hline\hline
\bfseries PARAMETER &
\bfseries AS GRAPH &
\bfseries INET--3.0 &
\bfseries FITNESS BA &
\bfseries BA MODEL \\
\hline
$N$, NUMBER OF NODES & 11461 & 11461 & 11461 & 11461\\
$L$, NUMBER OF LINKS & 32730 & 24171 & 34366 & 34366\\
$l(r_i\le 5\%,r_j)$ &  28602 & 22620 & 20929 & 15687 \\
$l(r_i\le 5\%,r_j\le 5\%)$ & 8919 & 3697 & 1426 & 1511 \\
MAXIMUM NODE DEGREE & 2432 & 2010 & 1793 & 329 \\
AVERAGE NODE DEGREE & 5.7 & 4.2 & 6.0 & 6.0\\
EXPONENT OF POWER--LAW DEGREE DISTRIBUTION & 2.22 & 2.22 & 2.255 & 3.0\\
\hline\hline
\multicolumn{5}{l}{$l(r_i\le 5\%,r_j)$ is the number of links connecting with the top 5\% rich nodes.}\\
\multicolumn{5}{l}{$l(r_i\le 5\%,r_j\le 5\%)$ is the number of
links connecting between the top 5\% rich nodes.}
\end{tabular}
\end{table*}

\subsection{AS Graph}
A number of studies \cite{Faloutsos99,Winick02,Albert00} on the
AS-level Internet topology used the so-called original AS
connectivity maps. The original maps are based on BGP routing
tables collected by the University of Oregon Route Views
Project~\cite{Oregon}. Chen~{\sl et~al} \cite{Chen02} constructed
the extended maps \cite{Michigan} using additional data sources,
such as the Internet Routing Registry (IRR) data and the Looking
Glass (LG) data. The extended maps have 20--50\% more links than
the original maps and provide a more complete picture of the AS
graph. The AS connectivity data used in this letter is an extended
map measured on May 26th 2001. Table~I shows some properties of
the data. For comparison, Table~I also shows the synthetic graphs
generated using the following network models.

\subsection{Barab\'asi-Albert Model}
The model \cite{Barabasi99a} generates networks with a power--law
degree distribution by using two generic mechanisms: {\sl growth},
where the network ``grows'' by attaching a new node with $m$ links
to $m$ different nodes present in the network; and {\sl
preferential attachment}, where new nodes are attached
preferentially to nodes that are already well connected. The
probability that a new node will be connected to node $i$ with
degree $k_i$ is

\begin{equation}
\Pi(i) = {k_i\over \sum_j k_j}.
\label{eq:barabasi}
\end{equation}

The BA model has generated great interest in various research
areas and has been used as a starting-point in research into the
error and attack tolerance of the Internet
\cite{Albert00,Albert00b,Holme02}. The properties shown in Table~I
are of a network where each new node has three new links ($m=3$)
which are preferentially connected to three already existing
nodes. Notice that this model generates networks with a power--law
link distribution of $P(k)\propto k^{-3}$ \cite{Barabasi99b}.

\subsection{Fitness BA Model}
The model \cite{Bianconi00} is a modification of the BA model. It
uses generalized preferential attachment which assures that, even
a relatively young node with a small number of links, can acquire
new links at a high rate if it has a large fitness parameter. The
reason we study this model is that, for the uniform fitness
parameter distribution, the network generated by this model has a
power--law exponent similar to that of the AS graph.

\subsection{Inet--3.0 Model}
The model \cite{Winick02} generates networks in three steps: 1)
build a spanning tree with all nodes that have degrees greater
than one, 2) connect all nodes with degree one to nodes in the
spanning tree with a linear preference, and 3) connect the
remaining free links in the spanning tree. The model was designed
to match the measurements of the original maps of the AS graph.
The number of links generated by the model depend on the number of
nodes and the exponent of the power--law. Choosing these two
parameters to match the AS graph of Table~I, the model typically
generates 26\% less links than the AS graph.

\section{The Rich--Club Phenomenon}

The rich--club is characterized by the rich--club connectivity and
the node--node link distribution, which measure the
interconnection between rich nodes.

\subsection{Rich--club connectivity}

Nodes in the network are sorted by decreasing number of links that
each node contains. There are instances where groups of nodes
contain identical numbers of links. Where this occurs, they are
arbitrarily assigned a position within that group. The node rank
$r$ denotes the position of a node on this ordered list. $r$ is
normalized by the total number of nodes $N$. The rich--club is
defined, for the purposes of this study, as nodes with rank less
than $r_{max}$ (e.g. 1\%). The rich--club connectivity $\phi(r)$
is defined as the ratio of the total actual number of links to the
maximum possible number of links between members of the
rich--club. The maximum possible number of links between $n$ nodes
is $n(n-1)/2$.

\begin{figure}[h]
\centerline{\psfig{figure=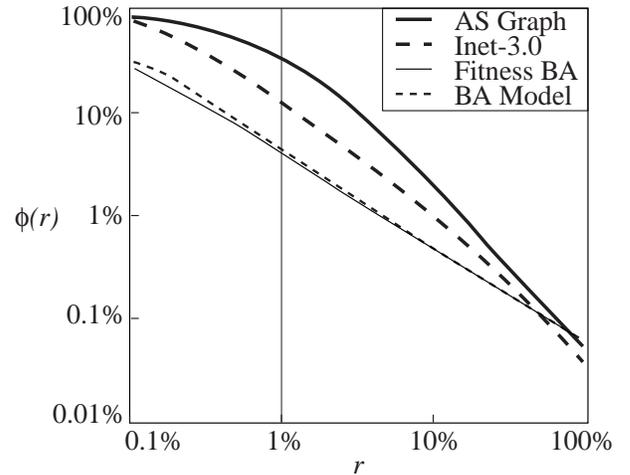,width=8cm}}
\label{fig:one} \caption{Rich--club connectivity $\phi(r)$ against
node rank $r$.}
\end{figure}

Figure~1 shows the rich--club coefficient $\phi(r)$ against node
rank $r$ on a log--log scale. It shows that the rich nodes of the
AS graph are very well connected between each other. The top 1\%
rich nodes have 32\% of the maximum possible number of links,
compared with $\phi(1\%)=18\%$ of the Inet-3.0  and only
$\phi(1\%)=5\%$ of the BA and the Fitness BA graphs.

\subsection{Node-node link distribution}

We define $l(r_i , r_j)$ as the number of links connecting nodes
with rank $r_i$ to nodes with rank $r_j$, where $r_i \le r_j$.
Figure~2 shows the node-node link distribution $l(r_i , r_j)$
against corresponding node rank $r_i$ and $r_j$. The  node ranks
are divided into 5\% bins.

\begin{figure}[h]
\centerline{\psfig{figure=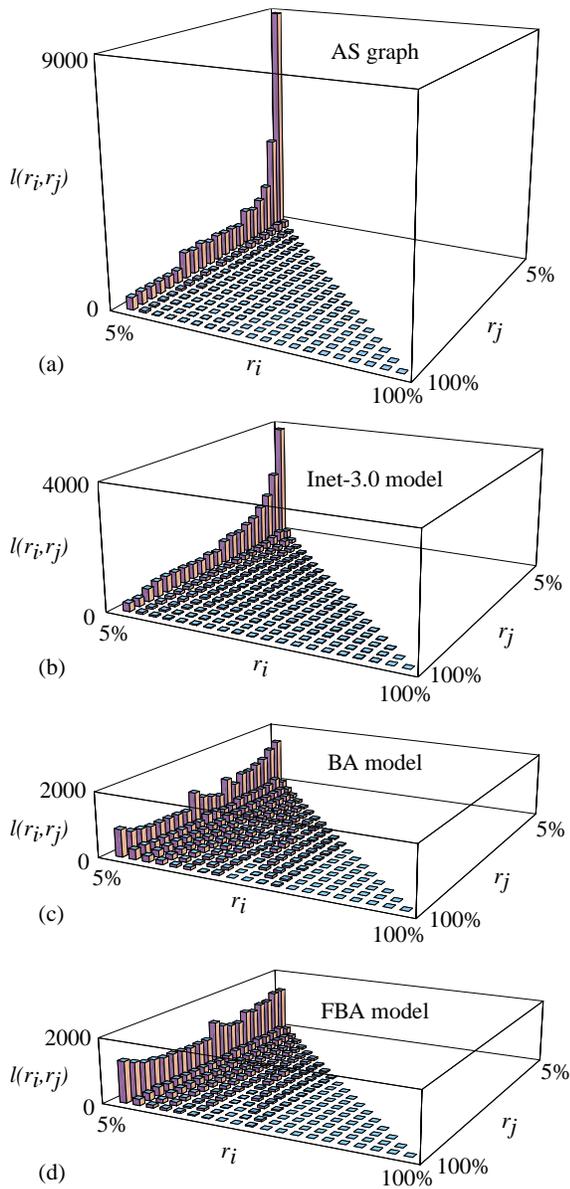,width=7.5cm}}
\label{fig:two} \caption{Node--node link distribution - number of
links $l(r_i,r_j)$ against node ranks $r_i$ and $r_j$.}
\end{figure}

In the AS graph (Fig.~2a), rich nodes are connected preferentially
to other rich nodes. The number of links between the top 5\% rich
nodes (far corner) is significantly larger than the numbers of
links connecting the rich nodes to other nodes with smaller
degrees (see the column with $r_i=5$\%).

The node-node link distribution of the Inet-3.0 (Fig.~2b) is
similar to that of the AS graph, however, the number of links
between the top 5\% rich nodes  of the Inet--3.0 model (far
corner, 3697 links) is significantly smaller than that of the AS
graph (8919 links).

The link distributions of the BA and the Fitness BA graphs
(Fig.~2c,~d) are different from that of the AS graph. The top 5\%
rich nodes of the BA and the Fitness BA graphs are connected to
all nodes with similar probabilities regardless of the node
degree. The graphs generated by these two models do not contain a
rich--club.

We end this section with the following observation: in the AS
graph, if we take the rich--club to comprise the nodes ranking 1\%
or less, the probability distribution of node degrees between the
members of the club is not a power--law, instead is more like the
distribution obtained from a random network.

\section{Discussion and Conclusion}

The AS graph has a core tier that we called the rich--club.
Membership of the rich--club, as defined above, could be further
limited by imposing the requirement of a relationship between the
members of the club and their connectivity, for example that the
average hop distance is less than $1.5$. This relationship can be
made more explicit if we approximate the rich--club to a random
network. In this case the average hop distance $\ell \approx
\ln(n)/\ln(<k>)=\ln(n)/\ln(\phi(r)(n-1)/2)$, where $n$ is the
number of nodes contained in the rich--club and $<k>$ is their
average node degree which is approximated by $\phi(r)(n-1)/2$.

We noticed that the BA and Fitness BA model do not have a
rich--club due to the growth dynamics of the models. All new links
connect with new nodes. Due to the preferential attachment, the
probability for a new node to become a rich node decreases as the
network grows. As a result, rich nodes are not well connected
between each other. This suggests a simple modification to these
models to generate a rich--club. As the network grows, new links
appear which are preferentially attached between the existing
nodes.

We believe that modeling the rich--club phenomenon is important
because the connectivity between rich nodes can be crucial for
network properties, such as network routing efficiency, redundancy
and robustness. In the AS graph, there is a large number of
alternative routing paths between the club members, their average
path length is very small (1 to 2 hops). The rich--club acts as a
super traffic hub and provides a large selection of shortcuts.
Hence scale-free models without the rich--club phenomenon may
under--estimate the efficiency and flexibility of the traffic
routing in the AS graph. Also, networks without the rich--club may
over--estimate the robustness of the network to a node  attack,
where the removal of a few of its richest club members can break
down the network integrity.

\bibliography{ZHOU_CL2003-0667}

\end{document}